\documentclass[10pt]{IEEEtran}
\IEEEoverridecommandlockouts
\usepackage{cite}
\usepackage{amsmath,amssymb,amsfonts}
\usepackage{algorithmic}
\usepackage{graphicx}
\usepackage{textcomp}
\usepackage{xcolor}
\usepackage{subfig}
\usepackage{comment}
\usepackage[left=0.625in,right=0.635in,top=0.75in,bottom=1.05in]{geometry}
\usepackage{multicol}
\def\BibTeX{{\rm B\kern-.05em{\sc i\kern-.025em b}\kern-.08em
    T\kern-.1667em\lower.7ex\hbox{E}\kern-.125emX}}
\begin{document}

\title{Quantifying Multimedia Streaming Quality: \\ 
A Practical Analysis using PIE and Flow Queue PIE
}

\author{\IEEEauthorblockN{Hemendra M. Naik, \\
Department of Computer Science and Engineering, \\
National Institute of Technology Karnataka, Surathkal, Mangalore - 575025, India \\
hemendranaik@gmail.com
}
\vspace{-1.2cm}
}

\maketitle

\begin{abstract}
The exponential growth of multimedia streaming services over the Internet emphasizes the increasing significance of ensuring a seamless and high-quality streaming experience for users. Dynamic Adaptive Streaming over HTTP (DASH) has emerged as a popular solution for delivering multimedia content over variable network conditions. However, challenges such as network congestion, intermittent packet losses, and varying network load continue to impact the Quality of Experience (QoE) perceived by the users. In this work, the main goal is to evaluate the effectiveness of using queue management and flow isolation techniques in terms of improving the overall QoE for DASH based multimedia streaming applications. Proportional Integral controller Enhanced (PIE) and Flow Queue PIE (FQ-PIE) are used as queue management and flow isolation mechanisms, respectively. The most distinctive aspect of this work is our assessment of QoE for multimedia streaming applications when multipath transport protocols, like Multipath TCP (MPTCP), are employed. Network Stack Tester (NeST), a Python based network emulator built on top of Linux network namespaces, has been used to perform the experiments. The parameters used for evaluating the QoE include bitrate, bitrate switches, throughput, Round Trip Time (RTT), and application buffer level. We observe that flow isolation techniques, combined with queue management and multipath transport, significantly improve the QoE for multimedia applications.
\end{abstract}


\section{Introduction}
\thispagestyle{empty}
In recent years, the usage of multimedia streaming platforms has significantly risen, resulting in increasing demands for low latency, minimal loss, and enhanced throughput. While end users may not directly articulate these performance needs, their tolerance for buffering in streaming applications remains notably low. Conversely, the substantial drop in memory prices has resulted in sizable unmanaged buffers across the Internet, causing considerable queuing delays, an issue commonly referred to as Bufferbloat \cite{Bufferbloat}. As a result, the necessity to effectively manage and control queues has become increasingly evident. Moreover, there has been notable enthusiasm surrounding the conceptualization and advancement of multipath transport protocols, including Multipath TCP (MPTCP) \cite{MPTCP-RFC-8684} and others, all aimed at achieving enhanced throughput and reliability.

The primary goal of this work is to empirically assess the effects of employing multipath transport protocols and queue management algorithms in conjunction with flow isolation techniques on multimedia streaming applications. To achieve this goal, we have considered a non-proprietary standard protocol proposed by Moving Pictures Experts Group (MPEG) called Dynamic Adaptive Streaming over HTTP (DASH) for multimedia streaming \cite{MPEG-DASH}, MPTCP for multipath transport, PI controller Enhanced (PIE) \cite{PIE} for queue management, and Flow Queue PIE (FQ-PIE) \cite{FQPIE} for flow isolation.

MPEG-DASH stands as a prominent adaptive bitrate streaming protocol extensively utilized in platforms such as YouTube and Netflix \cite{MPEG-DASH}. MPTCP, a standard protocol documented by the Internet Engineering Task Force (IETF) in RFC 8684, with most of its functionalities implemented in the Linux kernel, is adopted for multipath transport. PIE and FQ-PIE queue disciplines in the Linux kernel serve as queue management techniques, with and without flow isolation, respectively.

While previous experimental studies have individually examined the performance of MPTCP with queue management \cite{Cao2021}, MPTCP with MPEG-DASH \cite{Rakshit2023}, or even MPEG-DASH and queue management separately \cite{Kua2016}, no existing experimental evaluation comprehensively investigates all three technologies together, to the best of our knowledge. 

Network Stack Tester (NeST) \cite{NeST}, a Python based network emulator built on top of Linux network namespaces, has been chosen as the experimental platform. This choice is based on the unavailability of simultaneous support for MPEG-DASH, MPTCP, and PIE/FQ-PIE in existing simulators. NeST offers built-in APIs to emulate MPEG-DASH traffic, and configure MPTCP flows and PIE/FQ-PIE in the Linux kernel.

A critical aspect of this work involves collecting performance measurements of MPEG-DASH traffic while simultaneously accommodating bidirectional background flows. These background flows encompass various activities such as web traffic, Voice over Internet Protocol (VoIP) calls, and file transfers. Our primary emphasis lies in evaluating Quality of Experience (QoE) metrics, including bitrate, bitrate switches, throughput, Round-Trip Time (RTT), and application buffer level, specifically concerning MPEG-DASH traffic. While this study does not focus on fine-grained analysis of background traffic, we address the network's overall performance in relation to queuing delays. Our evaluations indicate that while PIE effectively manages queue delays within desired thresholds, FQ-PIE demonstrates superior overall QoE for MPEG-DASH streaming applications, owing to its flow isolation capabilities.

This paper makes the following three contributions: (i) performance evaluation of MPEG-DASH traffic using MPTCP and PIE/FQ-PIE in the presence of background flows such as web traffic, VoIP flows and file transfers, (ii) highlight the impact of flow isolation on multimedia streaming applications, and (iii) demonstrate how Linux network namespaces can be leveraged to perform near real-time experiments with state of the art network protocols within a controlled environment.

\section{Background and Related Work}
\subsection{Background}
\thispagestyle{empty}
\subsubsection{MPEG-DASH}
MPEG-DASH \cite{MPEG-DASH} is a widely adopted streaming protocol used for delivering multimedia content, such as video and audio, over the Internet. It has been developed to offer adaptive streaming, which automatically modifies the content's quality depending on the network conditions of the viewer. The multimedia content is broken up into manageable chunks and encoded at various quality levels in MPEG-DASH to provide several representations or versions. The MPD (Media Presentation Description) file, which serves as the manifest for the streaming session, is then used to make these representations accessible. The MPD file contains metadata, including the URLs of the available representations. The MPD file is requested from the server by the client device (such as a video player) during playback, which then chooses the best representation based on parameters like available bandwidth and device capabilities. The client then retrieves the content segments through HTTP, automatically switching between representations using adaptive algorithms to guarantee ongoing playback. MPEG-DASH offers versatility in content generation and delivery by supporting a wide range of codecs and media types. It is interoperable with common web servers and caches because it makes use of the already-existing HTTP infrastructure. This facilitates deployment and allows for simple integration with already-existing Content Delivery Networks (CDNs). Generally, MPEG-DASH offers a standardized method for adaptive streaming, improving the viewing experience by adjusting the video quality to changing network conditions, and enabling effective and scalable multimedia distribution over HTTP. MPEG-DASH is widely employed by major streaming platforms such as YouTube, Netflix, Prime Video, Disney+Hotstar, as well as by broadcasters globally, to deliver high-quality video content.

\subsubsection{MPTCP}
Most of the modern devices are multi-homed, featuring multiple network interfaces. For example, smartphones are typically equipped with at least two interfaces, enabling connections to both Wi-Fi and cellular networks. This multi-homing capability allows devices to communicate and access resources across multiple networks simultaneously, a concept also known as Wide Area Network (WAN) aggregation. Considering that TCP stands as the predominant protocol utilized over the Internet, MPTCP has emerged as the optimal choice for WAN aggregation due to its backward compatibility with TCP.

MPTCP data transfer makes use of the active subflows, each of which is confined within the current MPTCP session and looks like a standard TCP connection. This makes it possible for data to be transmitted effectively across any path that has enough capacity. RFC 6824 \cite{MPTCP-RFC-6824} outlines the core concepts of MPTCP, and subsequent RFCs, such RFC 8684 \cite{MPTCP-RFC-8684} refine specifics like security and deployment. Several TCP options like MP\_JOIN for dynamic path addition and MP\_CAPABLE for negotiation are introduced by these standards.

The three main stages of MPTCP include: connection setup, data transfer, and termination. The MP\_CAPABLE option is used during the TCP handshake to negotiate MPTCP capability during connection setup. The MP\_JOIN option allows pathways to be added dynamically in the future.

Three essential parts make up the architecture of MPTCP: the path manager, packet scheduler, and congestion control algorithm. The path manager is in charge of selecting which addresses to utilize, adding or eliminating new subflows, and controlling how they are advertised to peers. Data distribution is dynamically adjusted by the packet scheduler in response to subflow performance and delay. In the meantime, the congestion control algorithm maintains subflow fairness while identifying and reducing network congestion.

There are two openly available implementations of MPTCP for Linux: one categorized as out-of-tree and the other integrated upstream in Linux. We have used the upstream implementation of MPTCP in Linux 6.2.0-36-generic.

\subsubsection{PIE and FQ-PIE queue disciplines}
Active Queue Management (AQM) algorithms play a pivotal role in enhancing the performance and stability of modern computer networks by controlling queuing delays. These algorithms dynamically adjust queue lengths and drop or mark packets \cite{PIE} \cite{FQPIE} before queue delay increases significantly, thereby preventing network performance degradation and ensuring sufficient buffer space to accommodate intermittent bursts.

Proportional Integral controller Enhanced (PIE) \cite{PIE} is an enhancement to the PI controller, with a focus on achieving efficient queue control. It comprises three essential components: (i) calculation of per packet queue delay and, (ii) periodic calculation of drop/mark probability, and (iii) random dropping/marking during enqueue. The main goal of the PIE algorithm is to control the queue delay to a specified target value. The default value of target queue delay in PIE is 15 ms as specified in RFC 8033. However, it is a configurable knob and can be changed as per the requirements. 

Flow Queue PIE (FQ-PIE) \cite{FQPIE} offers a solution to resolve unfairness between responsive flows (e.g., TCP) and unresponsive flows (e.g., UDP) during network congestion. Algorithms like PIE effectively control the queuing delays by dropping or marking packets but often fall short in ensuring fair treatment among diverse flows sharing a common bottleneck. FQ-PIE emerges as a hybrid approach, combining flow isolation and PIE algorithm, specifically designed to address and rectify this challenge of unfairness in congested network scenarios.

In FQ-PIE, Flow Queuing (FQ) is implemented to assign each flow its dedicated queue, protecting responsive TCP flows from the impact of non-responsive ones, like multimedia traffic which exhibits a \emph{constant} bitrate. A hashing-based scheme is used, with flows hashed into buckets using the following five tuples: (source IP, destination IP, source port, destination port, protocol number). The Jenkins hash function is applied for this purpose in the Linux kernel.

A key objective of this work is to empirically investigate the efficacy of FQ-PIE's flow isolation feature in enhancing the overall QoE for MPEG-DASH multimedia applications.
To achieve this, we conduct experiments using both PIE and FQ-PIE and assess various performance metrics, including bitrate, bitrate switches, throughput, RTT, and application buffer level.
Additionally, we analyze the overall queue delay to gauge the effectiveness of PIE and FQ-PIE in terms of queue control. PIE and FQ-PIE are supported within the Linux kernel, and we utilize the same implementations for our evaluations.

\subsection{Related Work}

Mone et al. \cite{Mone2022} conducted a testbed-based analysis of Linux queue disciplines in a novel network topology under realistic application loads, including VoIP, HTTP, TCP, MPEG-DASH, UDP, and gaming traffic. They evaluated AQM mechanisms' performance over a 300-second experiment period. The study found that FQ-PIE demonstrated favorable results across various traffic types due to its burst allowance, flow queuing, and aggressive drop behavior. The authors suggest further exploration of FQ-PIE and its enhancements as potential alternatives to existing mechanisms like FQ-CoDel/CAKE. 

Kua et al. \cite{Kua2016} explore how emerging AQM schemes impact DASH content delivery, focusing on schemes like PIE, FQ-PIE, CoDel, and FQ-CoDel. They find that FQ-CoDel and PIE effectively safeguard DASH streams from interference and observe that the hybrid FQ-PIE scheme, which combines Flow Queue scheduling with PIE's burst tolerance, is the most promising for optimizing DASH performance as compared to PIE, CoDel, or FQ-CoDel. Their experiments, conducted on a TEACUP-based testbed, highlight the importance of AQM schemes in ensuring smooth multimedia streaming over varying network conditions.
	
Rakshit et al. \cite{Rakshit2023} conducted an empirical evaluation of MPEG-DASH traffic over MPTCP in comparison to traditional TCP in a mixed traffic environment. Utilizing Linux network namespaces and the state-of-the-art upstream implementation of MPTCP, they analyzed the impact of competing traffic on MPTCP's bandwidth utilization. Their study focused on QoE parameters such as bitrate switches, average throughput, application jitter, stalls frequency, and playback buffer level, alongside content parameters like segment lengths. However, the authors did not incorporate AQM algorithms in their evaluation. This paper advances upon their work by expanding the analysis to include AQM algorithms.
	
Cao et al. \cite{Cao2021} conducted a study comparing the performance of six AQM algorithms: Random Early Detection (RED), Fair Random Early Detection (FRED), Random Early Marking (REM), BLUE, and FQ under Low-Rate Denial-of-Service (LDDoS) attacks. They used a dual dumbbell topology along with MPTCP. They found that despite attacks on one path, MPTCP networks maintained functionality due to the robustness of alternative paths. FRED demonstrated superior throughput, while RED excelled in controlling queuing delay. However, the authors did not incorporate multimedia streaming traffic in their evaluation.

\section{Experimental Setup}
\thispagestyle{empty}
This section discusses the tools used for emulating various traffic types, experimental setup using Linux network namespaces and NeST, video dataset, and streaming configuration.

\subsection{Tools and Testbed}

We utilize Linux network namespaces and NeST to establish the experimental topology for this research. This method diverges slightly from those employed in prior studies, which often utilized physical testbeds or Virtual Machines (VMs) for experimentation. Linux network namespaces require notably fewer resources than VMs and offer cost-effectiveness compared to physical testbeds. NeST additionally streamlines the network emulation process through its intuitive Python APIs. Our experiments have been conducted on a system running Ubuntu 23.04 with kernel version 6.2.0-36-generic.

\begin{figure}[htbp]
\centerline{\includegraphics[width=\linewidth]{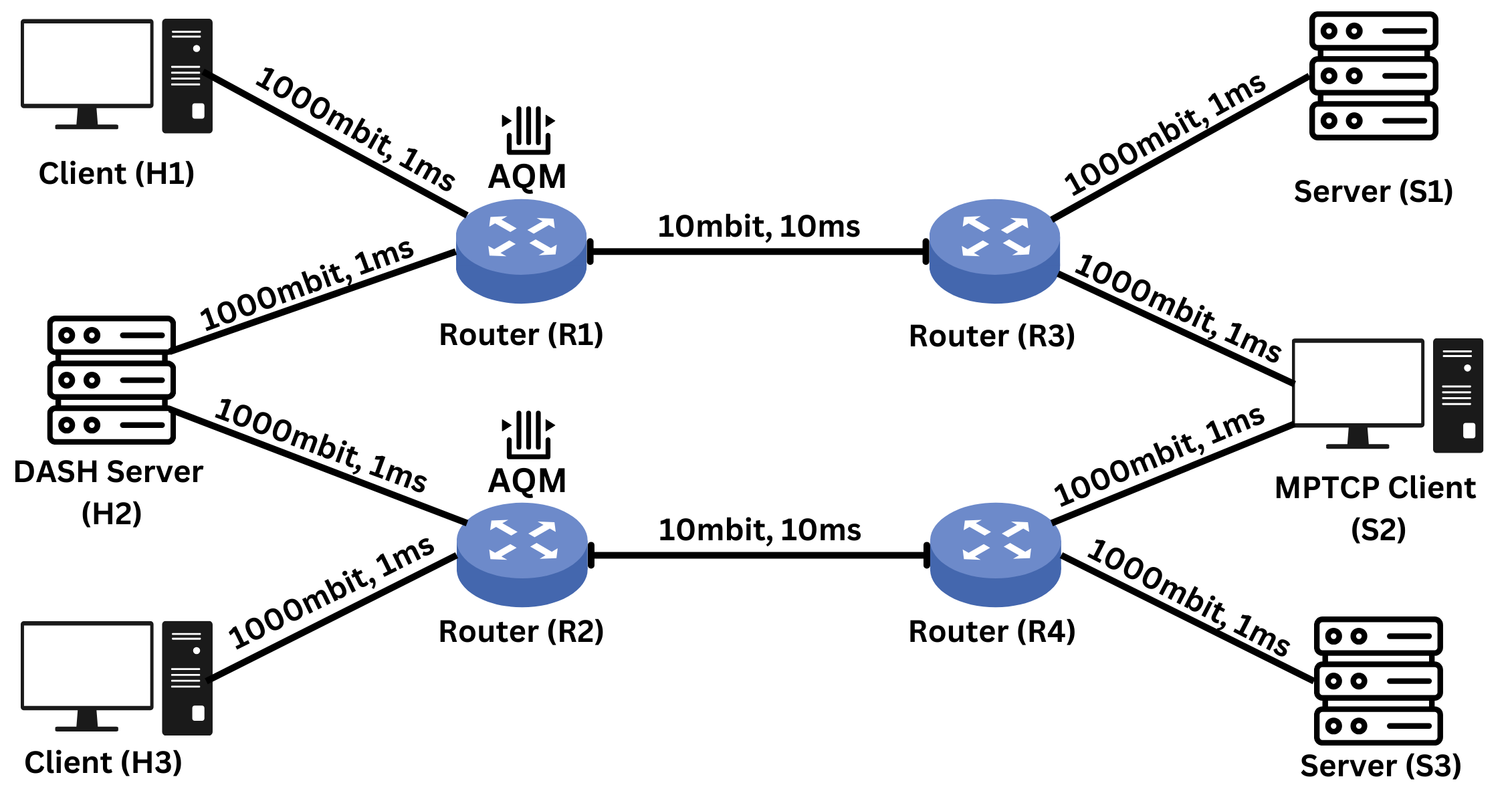}}
\vspace{-2mm}
\caption{Network Topology used for Experiments}
\label{Fig-Topology}
\vspace{-0.6cm}
\end{figure}

Our experiments use a topology used in \cite{Rakshit2023}, involving 2 multi-homed endpoints shown in Fig. \ref{Fig-Topology}. This topology emulates network congestion by orchestrating diverse traffic flows across bottleneck links (R1-R3 and R2-R4). To observe the functionality of PIE and FQ-PIE, we have modified the topology such that large amounts of payload flow from left to right. Thus, PIE and FQ-PIE are configured on the outgoing interface of R1 connecting it to R3 and the outgoing interface of R2 connecting it to R4 in respective experiments. 

\vspace{-2mm}

\subsection{Experimental Scenario}
In our experiments, only the MPEG-DASH stream has been configured to use MPTCP in anticipation to improve QoE. The fullmesh path manager along with uncoupled congestion control algorithm CUBIC has been set up between the MPTCP endpoints (H2 and S2). The DASH stream has been configured to flow simultaneously via two paths (H2-R1-R3-S2 and H2-R2-R4-S2). We used a simple HTTP Server in Python which adheres to the HTTP/1.1 specification. HTTP/1.1 ensures persistent connections which are beneficial for MPEG-DASH streaming. GPAC MP4 Client \cite{GPAC} has been used to stream the video. We incorporated FTP, VoIP, and HTTP traffic, three of the most common application traffic types on the Internet, into our testbed in an effort to emulate an actual Internet environment. The objective is to generate background traffic that accurately portrays the patterns of regular Internet usage, which include file transfers, video streaming, web surfing, and other applications that produce sporadic network traffic. To emulate FTP traffic, we utilized the iperf3 tool, a well known Linux network performance measurement utility. During our experiment, we established 5 FTP flows originating from host `H1' to server `S1' and 5 FTP flows from host `H3' to server `S3'. These flows persisted throughout the experiment, consistently transmitting data. For emulating VoIP traffic, SIPp, an open-source program, has been used, transmitting RTP data over UDP to replicate audio traffic. In our experimental setup, the SIPp client initiates 10 calls per second between each of the following: H1- S1 and H3-S3. Moreover, to emulate HTTP traffic, we utilized the httperf tool. HTTP GET requests were initiated from host `H1' to server `S1' and from host `H3' to server `S3', with a rate of 15 GET requests per second for each flow. Both flows persisted for a duration of 180 seconds. 


\begin{table}[htbp]
\caption{Traffic generated and the corresponding parameters.}
\begin{center}
\begin{tabular}{|p{0.8cm}|p{2cm}|p{1cm}|p{3cm}|}
\hline
\textbf{Traffic}  & \textbf{Tool Used and Version} & \textbf{Sender-Receiver pairs} & \textbf{Details}\\
\hline MPEG-DASH & Python v3.11.4 HTTP/1.1 Server and GPAC MP4 Client v2.0-rev2.0.0+dfsg1-2build1 \cite{GPAC} & H2-S2 & 1 MPEG-DASH stream between sender and receiver for 184 seconds. \\
\hline FTP (TCP) & iPerf v3.12 \cite{iperf3} & H1-S1 and H3-S3 & 5 FTP flows between each pair of sender and receiver. \\
\hline SIP (VoIP) & SIPp v3.6.0 \cite{sipp} & H1-S1 and H3-S3 & 10 VoIP calls initiated per second  \\
\hline HTTP & httperf v0.9.0 \cite{httperf} & H1-S1 and H3-S3 & 15 GET requests initiated per second for a 180 second duration. \\
\hline
\end{tabular}
\label{Table-Traffic}
\end{center}

\end{table}


\subsection{Choice of DASH dataset}

To evaluate MPEG-DASH based streaming applications, obtaining a dataset of pre-encoded chunks is essential. Two approaches exist for acquiring such data: manual encoding of videos into MPEG-DASH chunks or utilizing a standard open-source dataset. Opting for the latter option offers several advantages. Firstly, it saves significant time and effort, as pre-encoded datasets are readily available, eliminating the need for manual processing. Additionally, these datasets often undergo optimization by experts, resulting in enhanced video quality and reduced file sizes. Consequently, this optimization can lead to more efficient storage and expedited delivery over networks.

For our work, we used the MMSYS’22 dataset provided in  \cite{MMSYS22-Dataset}. Unlike previous works that relied on the MMSYS’12 \cite{MMSYS12-Dataset} dataset, we opted for the recently released MMSYS’22 dataset due to its comprehensive features. Notably, the MMSYS’22 dataset encompasses both audio and video streams, contrasting with the video-only content of the MMSYS’12 dataset. For our experiments, we specifically selected the ``Eldorado" video sequence, which spans 184 seconds in duration and is encoded using the AV1 codec, with a segment length of 4 seconds. The selection of a 4-second segment length was guided by insights from \cite{Rakshit2023} that indicated its suitability for improved streaming performance, particularly in terms of QoE metrics. These encoded video chunks feature 15 bitrate levels ranging from 145 Kbit (320 x 180 pixels) to 27500 Kbit (7680 x 4320 pixels). The audio chunks are encoded using the Advanced Audio Coding (AAC) codec with a single 128k bitrate track.

\subsection{Streaming Configuration}
In our experiments, we set up the GPAC player \cite{GPAC} to utilize the `grate' algorithm, which dynamically adjusts video quality based on estimated network capacity. Additionally, we configured GPAC to have a maximum buffer capacity of 10 seconds and a buffer threshold of 1 second for both audio and video streams individually. The MPTCP endpoints, H2 and S2 (in Figure \ref{Fig-Topology}), are configured with a `fullmesh' Path Manager and the `default' MPTCP scheduler integrated in Linux Kernel. We used the uncoupled congestion control algorithm supported by MPTCP's upstream implementation, ensuring distinct congestion windows for each flow. For each flow, we employed CUBIC, which is Linux's default congestion control algorithm.

    
\begin{table}[htbp]
\caption{Bitrate \& Resolution of AV1 encoded sequences}
\vspace{-2mm}
\begin{center}
\begin{tabular}{|c|c|c|c|}
\hline
\textbf{Bitrate (in bits)} & \textbf{Resolution} & \textbf{Bitrate (in bits)} & \textbf{Resolution} \\ \hline
145K & 320x180 & 5500K & 1920x1080 \\ \hline
180K & 384x216 & 7000K & 2560x1440 \\ \hline
350K & 512x288 & 11000K & 3840x2160 \\ \hline
500K & 640x360 & 15000K & 3840x2160 \\ \hline
700K & 768x432 & 20000K & 5120x2880 \\ \hline
900K & 1024x576 & 22500K & 7680x4320 \\ \hline
1400K & 1280x720 & 27500K & 7680x4320 \\ \hline
2750K & 1920x1080 &  &  \\ \hline
\end{tabular}
\label{tab1}
\end{center}
\vspace{-0.75cm}
\end{table}

\subsection{Configuration of PIE and FQ-PIE}
We have configured PIE and FQ-PIE algorithms on the intermediate routers R1 and R2 (in Figure \ref{Fig-Topology}) in the respective experiments. This aids in regulating queuing latency for traffic moving from R1 to R3 and R2 to R4, respectively. We evaluate the performance of these algorithms with three different values for the target queue delay in three different sets of experiments: 5ms, 10ms, and 15ms (recommended default in RFC 8033). Other parameters of PIE and FQ-PIE are unchanged. 

\section{Results and Analysis}
\thispagestyle{empty}

\begin{figure*}[h]
\centering
\captionsetup{font=footnotesize}
\begin{minipage}[t]{0.25\linewidth}
\includegraphics[width=\linewidth]{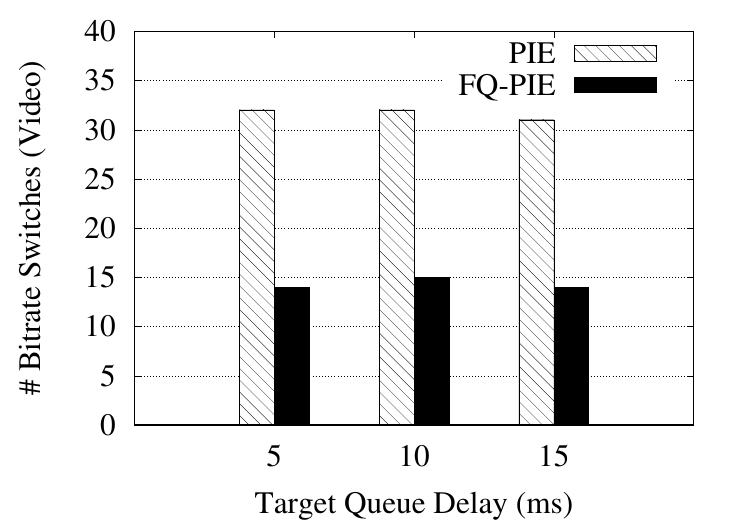}%
\caption{Bitrate Switches (Video)} \label{Fig-Video-Num-Bitrate}
\end{minipage}\hfill
\begin{minipage}[t]{0.25\linewidth}
\includegraphics[width=\linewidth]{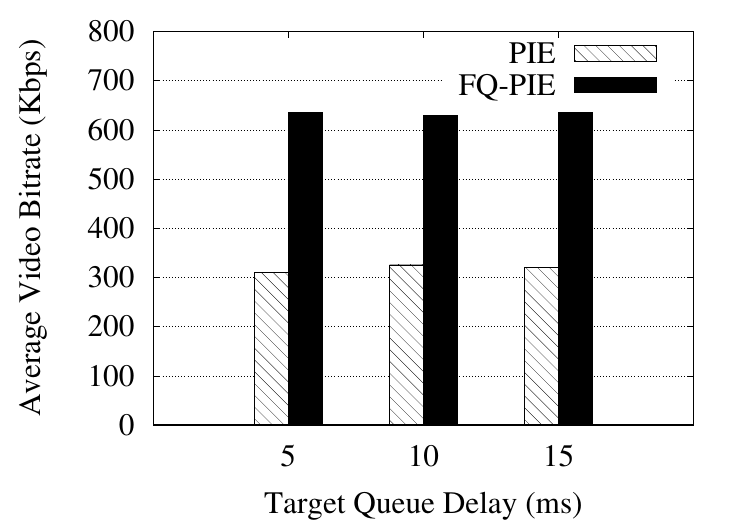}
\caption{Avg. Bitrate (Video)} \label{Fig-Video-Bitrate}
\end{minipage}\hfill
\begin{minipage}[t]{0.25\linewidth}
\includegraphics[width=\linewidth]{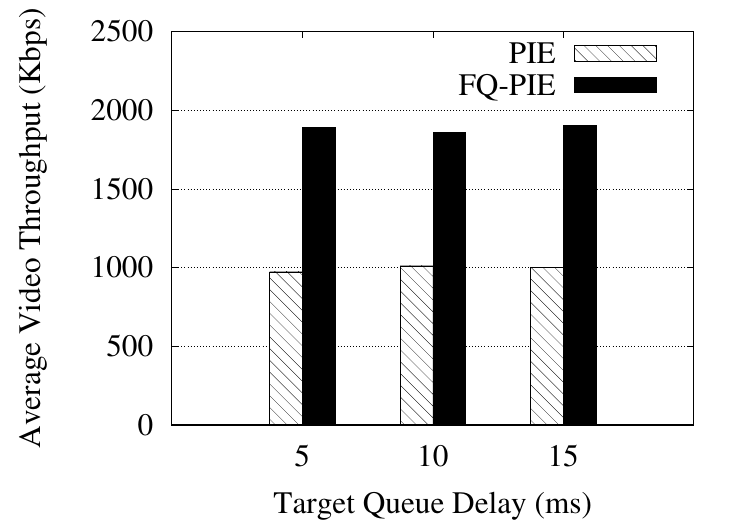}
\caption{Avg. Throughput (Video)} \label{Fig-Video-Throughput}
\end{minipage}\hfill
\begin{minipage}[t]{0.25\linewidth}
\includegraphics[width=\linewidth]{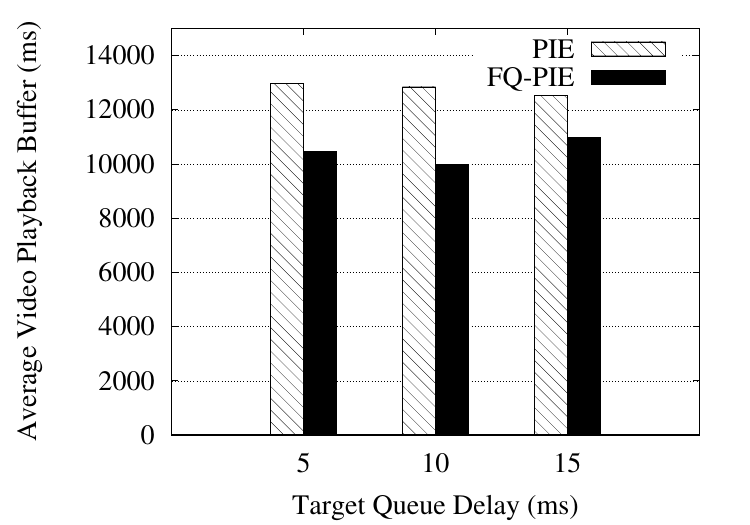}
\caption{Avg. Buffer (Video)} \label{Fig-Video-Buffer}
\end{minipage}\hfill

\vspace{0.15cm}

\begin{minipage}[t]{0.25\linewidth}
\includegraphics[width=\linewidth]{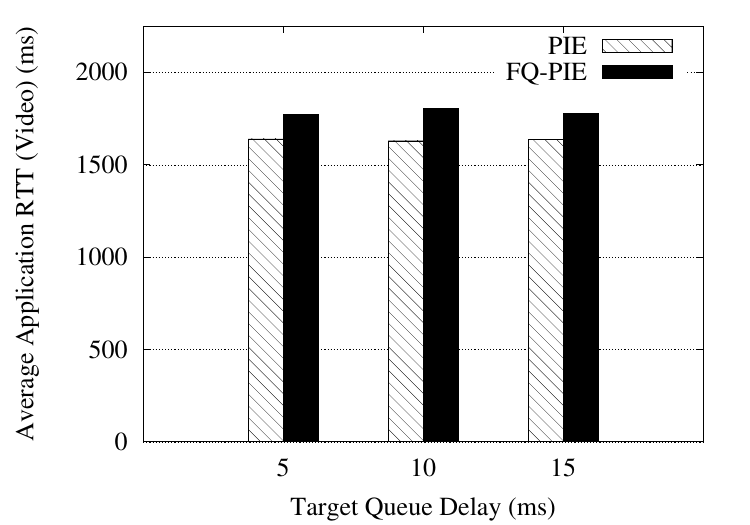}%
\caption{Avg. RTT (Video)} \label{Fig-Video-RTT}
\end{minipage}\hfill
\begin{minipage}[t]{0.25\linewidth}
\includegraphics[width=\linewidth]{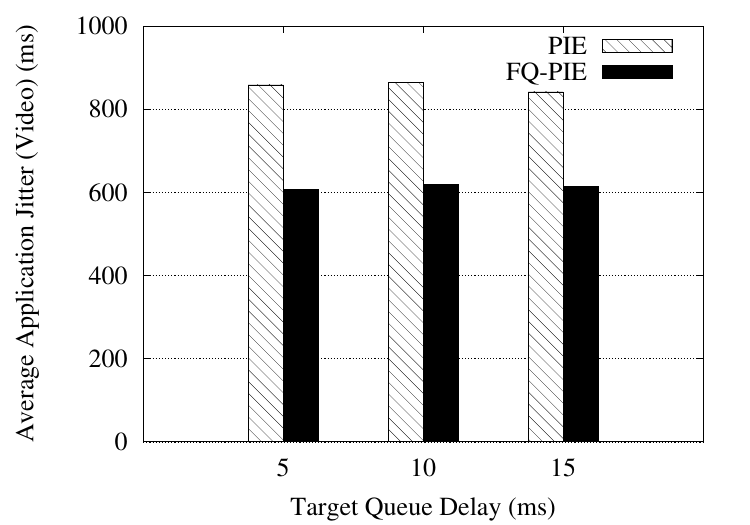}
\caption{Avg. Jitter (Video)} \label{Fig-Video-Jitter}
\end{minipage}\hfill
\begin{minipage}[t]{0.25\linewidth}
\includegraphics[width=\linewidth]{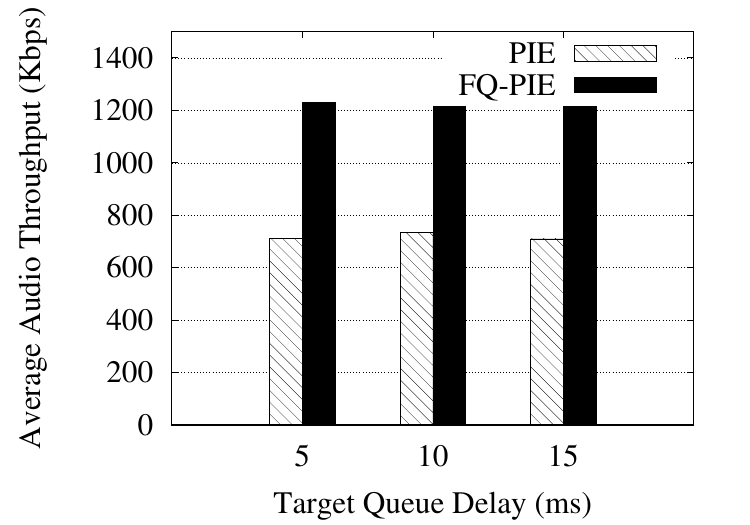}
\caption{Avg. Throughput (Audio)} \label{Fig-Audio-Throughput}
\end{minipage}\hfill
\begin{minipage}[t]{0.25\linewidth}
\includegraphics[width=\linewidth]{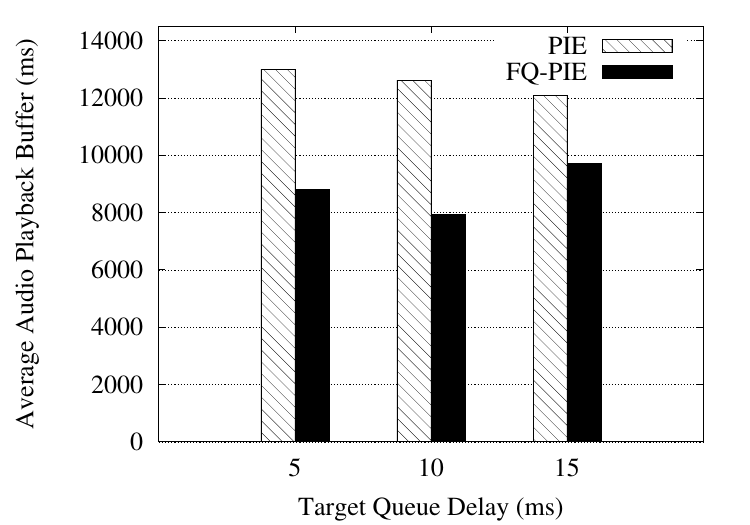}
\caption{Avg. Buffer (Audio)} \label{Fig-Audio-Buffer}
\end{minipage}\hfill

\vspace{0.15cm}

\begin{minipage}[t]{0.25\linewidth}
\includegraphics[width=\linewidth]{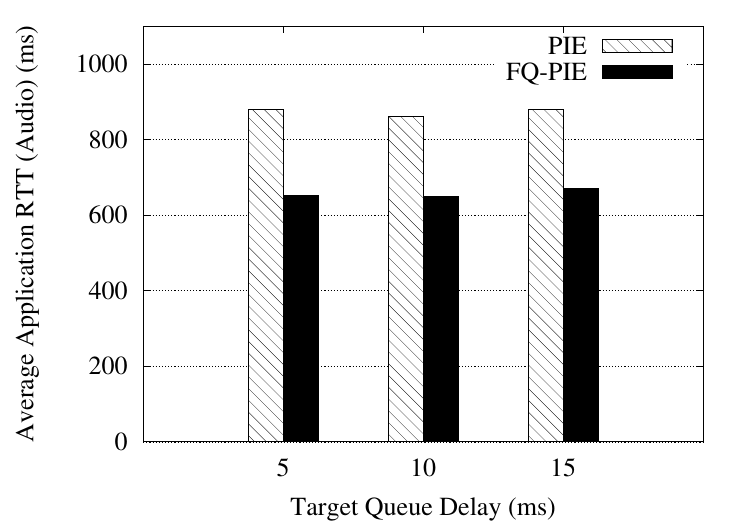}%
\caption{Avg. RTT (Audio)} \label{Fig-Audio-RTT}
\end{minipage}
\begin{minipage}[t]{0.25\linewidth}
\includegraphics[width=\linewidth]{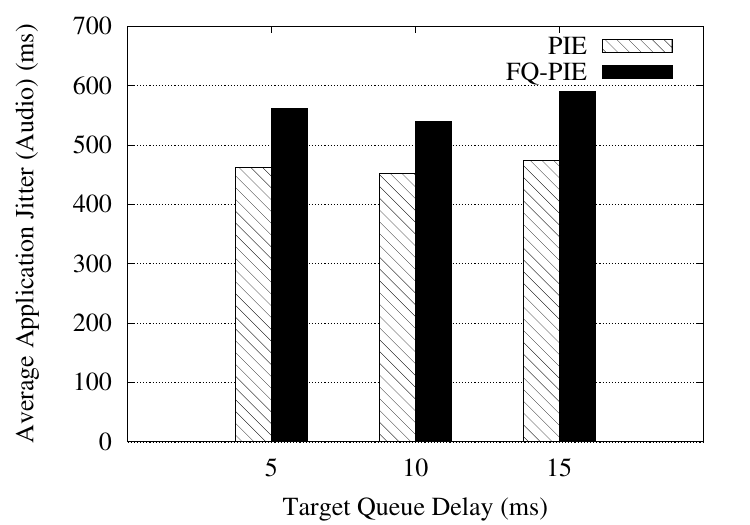}
\caption{Avg. Jitter (Audio)} \label{Fig-Audio-Jitter}
\end{minipage}
\vspace{-0.5cm}
\end{figure*}

This section presents the results obtained from the experiments and discusses the inferences that have been derived from them. To eliminate outliers and normalize the results, we have obtained 25 samples for every experiment. We consider various QoE metrics to assess the efficacy of DASH streaming, such as: (i) Number of Bitrate Switches, (ii) Average Bitrate, (iii) Average Throughput, (iv) Average Playback Buffer Level (v) Average Application RTT and (vi) Average Application Jitter. PIE and FQ-PIE are used as the underlying AQMs in the intermediate routers and the target queue delay in these AQM algorithms is varied as 5ms, 10ms and 15ms.

The \emph{Number of Bitrate Switches} refers to the number of times the streaming client switches between different bitrates during playback to accommodate changing network conditions. \emph{Average Video Bitrate} denotes the mean bitrate of the video stream over a specific duration, indicating the typical data rate of the video content. \emph{Average Throughput} denotes the average rate at which data is successfully transferred from the server to the client over the network during playback, providing an indication of the overall network performance experienced by the application. \emph{Average Playback Buffer} denotes the mean amount of data buffered by the streaming client during playback, representing the typical level of buffer occupancy throughout the streaming session. \emph{Average Application RTT} denotes the mean value of the total time elapsed between sending GET requests for MPEG-DASH chunks and receiving their responses from the server. Lastly, \emph{Average Application Jitter} refers to the mean value of variation in application RTT.

FQ-PIE's flow isolation technique ensures equitable bandwidth allocation among flows, reducing congestion-induced packet drops for MPEG-DASH streams. Consequently, this leads to a more stable observed bitrate (shown in Fig. \ref{Fig-Video-Bitrate}) and fewer bitrate switches (shown in Fig. \ref{Fig-Video-Num-Bitrate}) compared to PIE, thus improving the overall streaming experience. Moreover, FQ-PIE yields almost twice the throughput for audio and video streams compared to PIE as shown in Figures \ref{Fig-Video-Throughput} and \ref{Fig-Audio-Throughput}.

Additionally, it is important to note that audio statistics for average bitrate and bitrate switches were not collected due to the availability of a single bitrate representation for audio in MMSYS'22. With only one representation rate available, bitrate switches cannot occur. Therefore, the observed bitrate for audio streams is consistently close to 128 Kbps, which is the sole representation rate for audio in the dataset.

Yet another observation is that FQ-PIE causes the average playback buffer level (audio and video) (see Figures \ref{Fig-Video-Buffer} and \ref{Fig-Audio-Buffer}) of the DASH client to decrease as compared to that of PIE. This is due to the lack of flow differentiation in PIE that causes MPEG-DASH packets getting dropped. Combined with Head-of-Line (HoL) blocking, where DASH chunks cannot be consumed by the application until all corresponding packets arrive, playback buffer size tends to be slightly higher in case of PIE. The DASH client waits for retransmissions of dropped packets, leading to intermittent buffering during playback.
\thispagestyle{empty}

\begin{figure*}[h]
	\centering
	\begin{minipage}[t]{0.31\linewidth}
		\includegraphics[width=\linewidth]{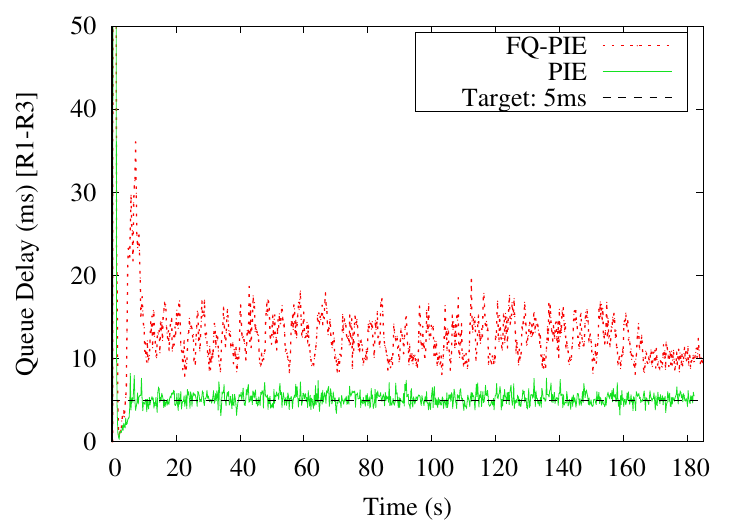}
	\end{minipage}\hfill
	\begin{minipage}[t]{0.31\linewidth}
		\includegraphics[width=\linewidth]{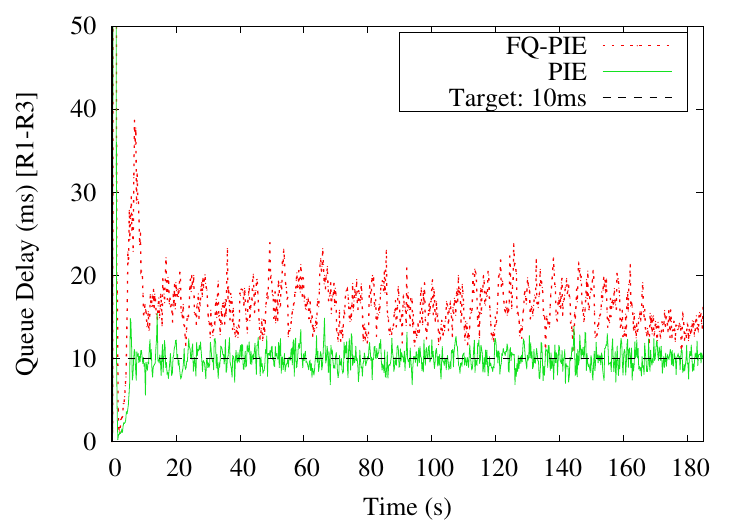}
	\end{minipage}\hfill
	\begin{minipage}[t]{0.31\linewidth}
		\includegraphics[width=\linewidth]{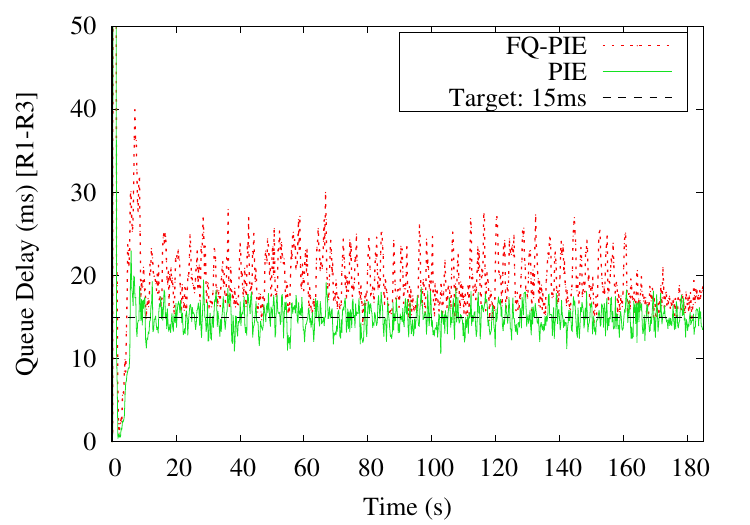}
	\end{minipage}\hfill
	\caption{Observed queuing delay on interface R1$\xrightarrow{}$R3, when target queue delay is set to 5ms, 10ms and 15ms} \label{Fig-QDelay-er1c}
	
	\vspace{0.2cm}
	
	\begin{minipage}[t]{0.31\linewidth}
		\includegraphics[width=\linewidth]{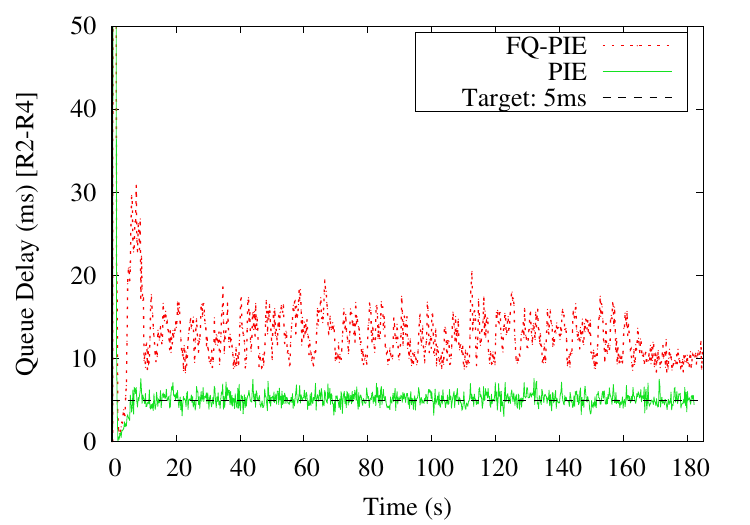}
	\end{minipage}\hfill
	\begin{minipage}[t]{0.31\linewidth}
		\includegraphics[width=\linewidth]{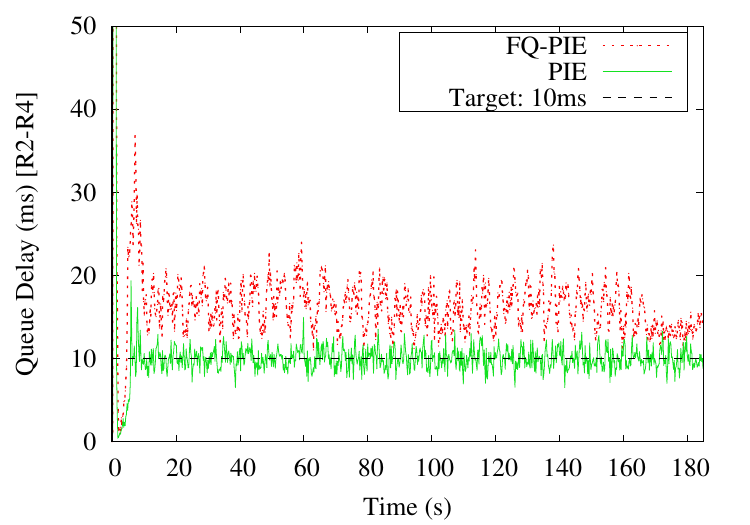}
	\end{minipage}\hfill
	\begin{minipage}[t]{0.31\linewidth}
		\includegraphics[width=\linewidth]{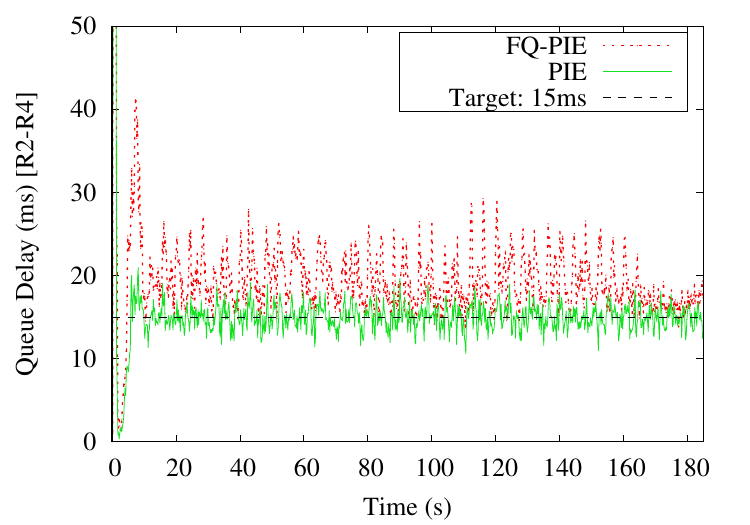}
	\end{minipage}\hfill
	\caption{Observed queuing delay on interface R2$\xrightarrow{}$R4, when target queue delay is set to 5ms, 10ms and 15ms} \label{Fig-QDelay-er2c}
	\vspace{-0.5cm}
\end{figure*}

As the bitrate of DASH video chunks increases to support higher video quality, resulting in larger data sizes, RTT increases accordingly (depicted in Figure \ref{Fig-Video-RTT}). Notably, FQ-PIE tends to stream higher quality video than PIE (600Kbps vs 300Kbps), yet its RTT (1700ms) is only slightly higher than PIE's (1600ms), indicating the effectiveness of FQ-PIE's optimization. In the MPEG-DASH dataset under consideration, there is only one bitrate representation (128Kbps) for the audio stream. Consequently, for downloading the identical set of audio chunks, FQ-PIE is anticipated to exhibit a shorter RTT  compared to PIE (depicted in Figure \ref{Fig-Audio-RTT}), owing to FQ-PIE's emphasis on flow isolation technique.

There are contrasting effects of FQ-PIE and PIE on the average application jitter for video and audio streams. Average application jitter for video streams reduces with FQ-PIE but it increases for audio streams (see Figures \ref{Fig-Video-Jitter} and \ref{Fig-Audio-Jitter}). This is because the dataset provides a \emph{fixed} bitrate for audio streams. When audio streams pass through a modified Deficit Round Robin (DRR) scheme in FQ-PIE during dequeue, it causes additional queue delays. However, the audio bitrate does not adapt because it is fixed. Nevertheless, the application buffer levels for audio with FQ-PIE are under proper control, and hence does not impact the user's perceived behavior.

In addition to measuring QoE metrics, we collected AQM statistics from the interfaces on the outgoing links between R1-R3 and R2-R4. To obtain these statistics for PIE and FQ-PIE AQMs, we utilized the tc (traffic control) utility from the iproute2 package in Linux. However, since the FQ-PIE AQM implementation in the Linux Kernel does not include average queue delay as part of the statistics, we added a new feature in its code to display all active PIE instances along with their statistics whenever tc is invoked to collect statistics. To calculate the average queuing delay of FQ-PIE at a given instant, we computed the mean queuing delay across all active PIE queues at that particular moment.

Figures \ref{Fig-QDelay-er1c} and \ref{Fig-QDelay-er2c} represent the plots of queue delay averaged out over 25 samples. It is evident that PIE is very effective in regulating the queue delay at the target value in all the scenarios. It can be observed that in spite of setting the target queue delay to 5ms, the observed queue delay oscillates from 10ms to 20ms in case of FQ-PIE. The oscillation in queue delay observed with FQ-PIE can be attributed to its management of multiple queues and employing a modified DRR algorithm to dequeue packets, each queue experiencing unique delays that are not synchronized. Consequently, variations in traffic intensity and packet arrival rates among these queues can lead to fluctuations in queue delay, resulting in deviations from the target value despite FQ-PIE's attempt to regulate it. Nevertheless, the queue delays with FQ-PIE are not significantly higher and trade-off quite well with the QoE results discussed earlier.

\section{Conclusion and Future Scope}
\thispagestyle{empty}
This paper discussed an experimental evaluation that combines three unique aspects: (i) multimedia applications (ii) using MPTCP as the underlying protocol and (iii) PIE and FQ-PIE as queue management algorithms. We used Linux network namespaces and NeST to perform network emulations. Our evaluations show that the flow isolation features of FQ-PIE combined with multipath resiliency provided by MPTCP leads to a significantly improved QoE for MPEG-DASH applications. Although queue control is slightly better with PIE, we believe using FQ-PIE improves the overall perception of the application behavior for the end users. As a next step, we aim to evaluate the performance of multimedia applications in extreme lossy environments while using MPTCP and FQ-PIE to understand the level of tolerance in multimedia applications when network environments are harsh.

\section*{Acknowledgments}

The author would like to express sincere gratitude to Dr. Mohit P. Tahiliani for his invaluable guidance and mentorship throughout this research work. The author would also like to thank Abhinaba Rakshit and Dayma Khan from the National Institute of Technology Karnataka, Surathkal, for their invaluable support and insights in navigating the intricacies of Multipath TCP (MPTCP) and MPEG-DASH. Their contributions were instrumental in shaping the research methodology and enhancing the quality of this work.

\bibliographystyle{IEEEtran}
\bibliography{IEEEabrv,References}
\thispagestyle{empty}
\end{document}